\documentclass[journal, 12pt, draftcls, onecolumn, twoside]{IEEEtran}

\usepackage{epsfig}
\usepackage{epstopdf}
\usepackage[utf8]{inputenc} 
\usepackage[T1]{fontenc}
\usepackage{url}
\usepackage{ifthen}
\usepackage[noadjust]{cite}
\usepackage[cmex10]{amsmath} 
\usepackage{amssymb}
\usepackage{graphicx}
\usepackage{subcaption}
\usepackage{mathtools}

\usepackage{multirow}
\usepackage{array}
\usepackage{makecell}
\usepackage{tikz}
\usepackage{diagbox}
\usepackage[export]{adjustbox}
\usepackage{caption, multirow, makecell}
\usepackage{algorithm}
\usepackage{algpseudocode}
\usepackage{breqn}

\usepackage{amsthm}
\usepackage{comment}
\usepackage{subcaption}

\usepackage{tabularx}
\usepackage{caption}
\usepackage{graphicx}
\usepackage{booktabs}
\makeatother

\usepackage[english]{babel}
\usepackage{lipsum}
\makeatletter
\renewcommand{\fnum@figure}{Fig. \thefigure}
\makeatother

\newtheorem{thm}{Theorem}
\newtheorem{exmp}{Example}
\newtheorem{cor}{Corollary}
\newtheorem{lem}{Lemma}
\newtheorem{rem}{Remark}

\begin{document}
	\title{An Optimal Two-Step Decoding at Receivers with Side Information in PSK-Modulated Index Coding} 
	
	\author{%
		\IEEEauthorblockN{ Navya Saxena, Anjana A. Mahesh, and B. Sundar Rajan\\}
		\IEEEauthorblockA{Department of Electrical Communication Engineering, Indian Institute of Science, Bengaluru \\
			E-mail: \{navyasaxena,anjanamahesh,bsrajan\}@iisc.ac.in}
	}
	\maketitle
\vspace{-1.5cm}
	\begin{abstract}
		This paper studies noisy index coding problems over single-input single-output broadcast channels. The codewords from a chosen index code of length $N$ are transmitted after $2^N$-PSK modulation over an AWGN channel. In ``Index Coded PSK Modulation for prioritized Receivers,'' \cite{divya} the authors showed that when a length-$N$ index code is transmitted as a $2^N$-PSK symbol, the ML decoder at a receiver decodes directly to the message bit rather than following the two-step decoding process of first demodulating the PSK symbol and equivalently the index-coded bits and then doing index-decoding. In this paper, we consider unprioritized receivers and follow the two-step decoding process at the receivers. After estimating the PSK symbol using an ML decoder, at  a receiver, there might be more than one decoding strategy, i.e., a linear combination of index-coded bits and different subsets of side information bits, that can be used to estimate the requested message. Thomas et al. in [``Single Uniprior Index Coding With Min–Max Probability of Error Over Fading Channels,''] \cite{anoop} showed that for binary-modulated index code transmissions, minimizing the number of transmissions used to decode a requested message is equivalent to minimizing the probability of error. This paper shows that this is no longer the case while employing multi-level modulations. Further, we consider that the side information available to each receiver is also noisy and derive an expression for the probability that a requested message bit is estimated erroneously at a receiver. We also show that the criterion for choosing a decoding strategy that gives the best probability of error performance at a receiver changes with the signal-to-noise ratio at which the side information is broadcast. Hence, for a given index coding problem and a chosen index code, we give an algorithm to choose the best decoding strategy at the receivers. The above results are shown to be valid over fading channels also.  
	\end{abstract}

	\begin{IEEEkeywords}
		Broadcast channels, Noisy Index Coding, Side information.
	\end{IEEEkeywords}

	\IEEEpeerreviewmaketitle
	\section{Introduction}
	The Index Coding Problem (ICP), introduced in \cite{birk}, is now a well-studied problem in network information theory which aims to characterize the optimal communication rates and  coding schemes for broadcasting multiple messages to a system of receivers with side information. Instances of ICP arise in satellite communications \cite{ongho}, cache-aided content broadcasting \cite{MAN}, coded computing \cite{LMYA}, etc. It consists of a central server that has access to a set $\mathcal{X}$ of messages, broadcasting to a group of receivers, each of which knows a subset of the messages in $\mathcal{X}$ \emph{a priori} as side information and requests another subset of messages from the server. The solution of an ICP, which is a set of server transmissions that satisfy all the receivers, is called an index code, and the number of transmissions in it is called its length. If all the transmissions in an index code are linear combinations of the messages in $\mathcal{X}$, then it is called linear. If each coded transmission is formed using a single generation of the messages in $\mathcal{X}$, the index code is said to be scalar. An optimal index code is one with the minimum possible number of transmissions. 	
	
	Bar-Yossef et al. \cite{yossef} studied a particular type of ICP in which each receiver demands a single unique message,  which was represented using a directed graph called a side information graph. In \cite{yossef}, it was proved that for a given ICP, the length of an optimal scalar linear index code is equal to a graph functional called minrank of the corresponding side information graph. Ong and Ho \cite{ongho} classified ICPs depending on the nature of demands and side information of receivers. If the side information at each receiver is unique, then it is called a uniprior ICP. If the demand of every receiver is unique, then it is called a unicast ICP. Further, in a unicast ICP, if each receiver demands only a single message, it is called a  single unicast ICP.  General ICPs that are neither uniprior nor unicast are termed multicast/multiprior. 
	
	Most of the literature concerning ICPs considers server transmissions over noiseless broadcast channels, while in practice, the transmissions can never be noise-free. Noisy ICPs have been studied in \cite{natrajan}\textendash\cite{divya}, among others. Multi-level modulations have been used for the transmission of index codewords in some of these works. A special case of ICP over AWGN channel with quadrature amplitude modulation has been studied in \cite{natrajan}. Phase shift keying was the chosen modulation scheme for transmitting index-coded bits over the AWGN channel in \cite{anjana} as well as \cite{divya}, which was called index-coded PSK modulation. In these papers, for a chosen binary index code of length $N$, the $N$ index-coded bits are mapped to a signal point in $2^N$- PSK constellation. 
	
	There might be situations in which the server gives higher priority to some of the receivers, which may be based on the premium paid. Such a prioritized receiver system is considered in \cite{divya}, \cite{TSR}. The techniques used in these papers are aimed at giving the best probability of error performance of the highest priority receiver, but this might lead to the performance of lower priority receivers deteriorating considerably.  However, there might also be cases where the server does not prioritize any of the receivers. For instance, a service provider (say, a television channel) may serve all its users equally as all have paid equal money. In this work, we consider noisy index coding problems with unprioritized receivers over AWGN broadcast channels. 
	
	In the first half of this paper, we assume that the side information available to the receivers and noise-free, whereas, in the second half, we assume that the side information at the receivers is obtained from binary modulated broadcast transmissions by the server, maybe, during an earlier off-peak window, and hence are noisy. Due to differences in receiver sensitivities or detection thresholds, different subsets of the messages might not be decoded, i.e., get erased, at different receivers resulting in non-identical subsets of the message set $\mathcal{X}$ as side information at different receivers. 
	
	Decoding a requested message at a receiver involves the two-step process of first performing ML decoding to estimate the transmitted PSK symbol, and hence the index-coded bits, and then decoding the requested message bit by using some linear combination of the estimated index-coded bits and its side information. At any receiver, there might be several linear combinations of the index-coded bits, which, along with its side information, could be used for decoding a particular requested message, which we call possible decoding strategies at that receiver. For binary-modulated transmissions, it was shown in \cite{anoop} that the best decoding strategy w.r.t probability of error is a linear combination of the minimum number of index-coded bits. In this paper, we show that when the index code is transmitted using multi-level modulation, minimizing the number of index-coded bits used in index-decoding post maximum-likelihood (ML) decoding of the PSK symbol need not result in the best probability of error performance at the receivers. 
	
	For $M$-PSK modulated transmission of index codes over an AWGN channel, at a receiver which performs the two-step decoding process as described above, we derive a criterion for selecting a decoding strategy that results in the best probability of error performance. There might be other two-step decoding processes, that use a non-ML decoder for estimation of PSK symbol and a different decoding strategy than the one shown to be optimal in our setting, that may result in a better probability of error performance at a receiver by virtue of error cancellations. Hence, the optimality of the decoding strategy w.r.t minimizing the probability of error at a receiver is only among the two-step decoding schemes employing ML estimation of PSK symbol in the first step. The main contributions of this work are listed below:
	\begin{itemize}
		\item We prove that for index code transmission using multi-level modulation, the probability of error performance at a receiver does not depend on the number of index-coded bits used in decoding a requested message.
		\item Complete theoretical analysis is carried out for index code transmission over an AWGN channel using multi-level modulation by deriving the expression for the probability of bit error obtained by any decoding strategy at a receiver when the side information at it is a) noiseless, and b) noisy. 
		\item For selecting an optimal decoding strategy with respect to the probability of error performance at a receiver having noisy side information, different criteria are derived for high and low values of the SNR at which the side information messages are broadcast.
		\item Based on the above criteria, for a given mapping of index-coded bits to the PSK constellation, an algorithm is presented that outputs the best decoding strategy for a requested message at a given receiver. 
		\item Simulation results validating that the decoding strategy chosen based on the criteria in this paper gives the best probability of error performance at the receivers are also provided. 
		\item Assuming perfect channel state information at each receiver, we establish that the proposed results remain valid even when the broadcast channel between the source and the receivers is a fading channel.

	\end{itemize}
	\subsection{Organization and Notation}
	The rest of this paper is organized as follows. The paper is divided into two sections,  Section \ref{noiseless_si} which studies ICPs with noise-free side information at the receivers, and Section \ref{noisy_si}, where the side information at the receivers are assumed to be noisy. In Section \ref{noiseless_si}, the system model is described in subsection \ref{model1}. The expression for probability of decoded message error at a receiver with noiseless information is derived, and an algorithm for finding an optimal decoding strategy is presented  in the following subsection \ref{mainresult1}. The results in subsection \ref{mainresult1} are explained using a detailed example in the final subsection \ref{sec:example1} of section \ref{noiseless_si}. Similarly, in Section \ref{noisy_si}, which considers ICPs with noisy side information at the receivers, the system model is discussed in subsection \ref{model2} followed by main results in subsection \ref{mainresult2} and an illustrative example in subsection \ref{sec2:example}. The validity of the results in this paper over fading channels is established  in Section \ref{fading_analysis}, and simulation results are presented in Section \ref{simulationexample}. Finally, the paper is concluded by giving a summary of the contributions in Section \ref{conclusion}.

	The mathematical notations used in the paper are as follows: The binary field consisting of the elements $0$ and $1$ is denoted as $\mathbb{F}_2$. The set $\{1,2,3,...,n\} $ is denoted by $[n]$. $f(y)$ denotes any function $f$ which takes input argument $y$. A vector is represented by a lowercase bold-face letter, as in $\textbf{x}$, while a matrix is represented by an upper-case bold-face letter, as in $\textbf{L}$. $x_i$ represents the $i^{th}$ component of $\textbf{x}$, while $\textbf{x}_B$ denotes the vector defined as $\textbf{x}_B=(x_i : i\in[m], i\in[B])$. For a matrix $\mathbf{A}$,  $\mathbf{A}^T$ denotes its transpose. The symbol $\oplus$ is used to denote the XOR of the operands. For a set $S$ consisting of $m$ elements, $S(i)$ denotes the $i^{\text{th}}$ element in $S$, for $i \in [m]$.
	
	\section{Receivers with Noiseless Side Information} 
	\label{noiseless_si}
		\subsection{System Model}
	\label{model1}
	
	We consider an ICP with $m$ messages denoted by $\mathcal{X}=\{x_1,x_2,\ldots x_m\}$ where $x_i\in\ \mathbb{F}_2$ and $n$ receivers denoted by $\mathcal{R}=\{R_1,R_2,\ldots R_n\}$. Receiver $R_i$ has side information denoted by ${K_i} \subseteq \mathcal{X}$ and demands another subset $W_i \subseteq  \mathcal{X} \setminus K_i $, $\forall{i}\in[n]$.  Without loss of generality, we assume that $|W_i|  = 1$, $\forall i \in [n]$ since if a receiver $R_i$ wants more than one message, then ${R_i}$ can be split into several receivers each wanting a single message and having $K_i$ as side information.  For a given ICP, an index code of length $N$ consists of:
	\begin{enumerate}
		\item an encoding scheme, $\mathcal{E} : \mathcal{X} \rightarrow \mathbb{F}_2^N$ and
		\item a set of $n$ decoding functions, $\{\mathcal{D}^i\}_{i \in [n]}$, $\mathcal{D}^i : \mathbb{F}_2^N \times \mathbb{F}_2^{|K_i|} \rightarrow \mathbb{F}_2^{|W_i|}$ s.t, for $\mathbf{x} \in \mathbb{F}_2^n$, $\mathcal{D}^i(\mathcal{E }(\mathbf{x}),K_i) = W_i$. 
	\end{enumerate}
	
	We consider the encoding scheme to be linear, which can be represented using an $n \times N$ matrix $\mathbf{L}$ over $\mathbb{F}_2$ and the index-coded vector is represented as $\mathbf{y} = (y_1, y_2, \cdots, y_N)$. We assume that the index-coded vector is transmitted after modulating it using a $2^N$- PSK constellation. Let the mapping from $N$ index-coded bits to the $2^N$- PSK signal set be denoted as $\mathcal{M}$. At receiver $R_i$, the received signal is the complex number $c_i = \mathcal{M}(\mathbf{y}) +n_i$, where $n_i$ is the white Gaussian noise, from which $R_i$ estimates the transmitted PSK signal point or equivalently the index-coded bits by performing minimum Euclidean distance decoding. The vector of estimated index-coded bits at $R_i$ is denoted as $\hat{\mathbf{y}}^i = (\hat{y}^i_1, \hat{y}^i_2, \cdots , \hat{y}^i_N)$.

	For a linear encoding scheme, the decoding function $\mathcal{D}^i$ at receiver $R_i$ is a linear combination of all or a subset of the index-coded bits and a subset of its side information. At any receiver, there could be multiple linear combinations of index-coded bits and side information that could give its requested message, which we call the possible decoding strategies at that receiver. At receiver $R_i$, if there are $r$ different decoding strategies, they are denoted as $\mathcal{D}^i_1, \mathcal{D}^i_2, \cdots, \mathcal{D}^i_r$, each of which is a function of the estimated index-coded bits in $\hat{\mathbf{y}}^i$ as well as its side information. The following example illustrates the notations defined above.

\begin{figure}
		\centering	
		\captionsetup{justification = centering}
		\includegraphics[scale=0.6]{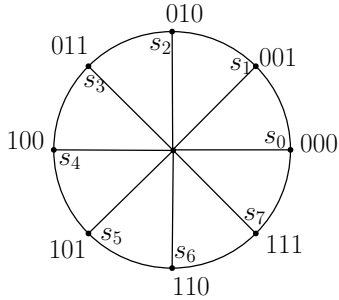}
		\caption{$\mathcal{M}_1$-$8$-PSK Mapping used in Example \ref{example-1}.}
		\label{8-psk}	
	\end{figure}

	\begin{exmp}
		\label{example-1}
		Consider an ICP with  $m=n=5$ and $W_i=x_i, \ \forall{i} \in \{1,2,3,4,5\}$. The side information at the receivers are ${K_1=\{x_2,x_3,x_4,x_5\}}$,  $K_2=\{x_1,x_3,x_4,x_5\}$, $K_3=\{x_2,x_4\}$,  $K_4=\{x_1\}$, and $K_5=\{x_3\}$.
		For this ICP, the length of an optimal linear index code is $N=3$.  We consider the linear encoding scheme specified by 
		$\mathbf{L} = \small\begin{bmatrix}
			1 & 0  & 0 & 1 & 1 \\
			1 & 1 & 1 & 1 & 1 \\
			1 & 0 & 0 & 1 & 0
		\end{bmatrix}^T$ and the index-coded vector is $\mathbf{y}=(y_1,y_2,y_3)$, where $\mathbf{y}=\mathbf{x}\mathbf{L}$, for $\mathbf{x} \in \mathbb{F}_2^5$. We transmit the above index code using the 8-PSK mapping given in Fig. \ref{8-psk} through an AWGN channel. At receiver $R_i$, after performing ML decoding on the received signal, the vector of estimated index-coded bits is $\mathbf{\hat{y}}^i= (\hat{y}^i_1, \hat{y}^i_2,\hat{y}^i_3)$. The different decoding strategies at each of the receivers are given below. 
		
	At $R_1$:
	\begin{itemize}
		\item[]$\hat{x}_1=\mathcal{D}_1^1(\mathbf{\hat{y}}^1,K_1) := \hat{y}^1_1  \oplus x_4  \oplus x_5$
		\item[]$\hat{x}_1=\mathcal{D}_2^1(\mathbf{\hat{y}}^1,K_1) :=\hat{y}^1_2 \oplus x_2 \oplus x_3 \oplus x_4 \oplus x_5$
		\item[]$\hat{x}_1=\mathcal{D}_3^1(\mathbf{\hat{y}}^1,K_1) :=\hat{y}^1_3 \oplus x_4$
		\item[]$\hat{x}_1=\mathcal{D}_4^1(\mathbf{\hat{y}}^1,K_1) :=\hat{y}^1_1 \oplus \hat{y}^1_2 \oplus \hat{y}^1_3 \oplus x_4 \oplus x_2 \oplus x_3$
	\end{itemize}
	
	At $R_2$:
	\begin{itemize}
		\item[]$\hat{x}_2=\mathcal{D}_1^2(\mathbf{\hat{y}}^2,K_2) :=\hat{y}^2_1 \oplus \hat{y}^2_2 \oplus x_3$
		\item[]$\hat{x}_2=\mathcal{D}_2^2(\mathbf{\hat{y}}^2,K_2) :=\hat{y}^2_2 \oplus x_1 \oplus x_3 \oplus x_4 \oplus x_5$
		\item[]$\hat{x}_2=\mathcal{D}_3^2(\mathbf{\hat{y}}^2,K_2) :=\hat{y}^2_2 \oplus \hat{y}^2_3 \oplus x_3 \oplus x_5$
		\item[]$\hat{x}_2=\mathcal{D}_4^2(\mathbf{\hat{y}}^2,K_2) :=\hat{y}^2_1 \oplus \hat{y}^2_2 \oplus \hat{y}^2_3 \oplus x_1 \oplus x_4 \oplus x_3$
	\end{itemize}
	
	At $R_3$:
	\begin{itemize}
		\item[]$\hat{x}_3=\mathcal{D}_1^3(\mathbf{\hat{y}}^3,K_3) :=\hat{y}^3_1 \oplus \hat{y}^3_2 \oplus x_2$
	\end{itemize}
	
	At $R_4$:
	\begin{itemize}
		\item[]$\hat{x}_4=\mathcal{D}_1^4(\mathbf{\hat{y}}^4,K_4) :=\hat{y}^4_3 \oplus x_1$
	\end{itemize}
	
	At $R_5$:
	\begin{itemize}
		\item[]$\hat{x}_5=\mathcal{D}_1^5(\mathbf{\hat{y}}^5,K_5) :=\hat{y}^5_1 \oplus \hat{y}^5_3$.
	\end{itemize}
	\end{exmp}
	
	From the above example, we saw that for a given index code, there could exist different ways in which a receiver can decode its requested message.  When multi-level modulation schemes (here, PSK) are used to transmit index-coded bits, the event of each bit going wrong is not independent of each other. Hence, the probability of error performance of each index-coded bit is not the same. In the following section, we prove that, for transmission employing multi-level modulations, the probability of error in decoding a requested message does not depend on the number of index-coded bits used in decoding it and derive a criterion for choosing an optimal decoding strategy at a receiver. 
	
	\subsection{Main Results} 
	\label{mainresult1}
	
	For a chosen index code of length $N$ of a given ICP and a chosen mapping $\mathcal{M}$ of index codewords to PSK signal points, let the $2^N$- PSK signal set configuration used for transmission be denoted as $\mathcal{S}_{2^N,\mathcal{M}}$. For example, with $N=3$ and $\mathcal{M}$ as the decimal to binary mapping, $\mathcal{S}_{8,\mathcal{M}}$ is shown in Fig. \ref{8-psk}. At a receiver $R_i$, denote the set $\{\hat{y}^i_j\}_{j \in [N]}$ of estimated index-coded bits by $\hat{Y}^i$. For a decoding strategy $\mathcal{D}_j^i$ at $R_i$, let the indices of the index-coded bits used by $\mathcal{D}_j^i$ be denoted as $I(\mathcal{D}_j^i)$. Hence, from the set $\hat{Y}^i$ of estimated index-coded bits, $\mathcal{D}_j^i$ uses a linear combination of the subset $\hat{Y}^i_{I(\mathcal{D}_j^i)} \subseteq \hat{Y}^i$  of estimated index-coded bits along with the side information $K_i$ to decode $W_i$. The XOR of the bits in $\hat{Y}^i_{I(\mathcal{D}_j^i)}$ corresponding to a signal point $s_k \in \mathcal{S}_{2^N,\mathcal{M}}$ is denoted as $\bigoplus\limits_{s_k} \hat{Y}^i_{I(\mathcal{D}_j^i)}$. 
	
	For a pair of signal points $s_j, s_k \in \mathcal{S}_{2^N,\mathcal{M}}$, the distance between them in $\mathcal{S}_{2^N,\mathcal{M}}$ is denoted as $dist(s_j,s_k)$. The minimum Euclidean distance of an  $M$-PSK signal set is denoted as $\Delta_{M-\text{PSK}}$. For a decoding strategy $\mathcal{D}_j^i$ and a chosen $2^N$- PSK signal set configuration $\mathcal{S}_{2^N,\mathcal{M}}$, the set of signal pairs $(s_j,s_k)$, such that $s_j,s_k \in \mathcal{S}_{2^N,\mathcal{M}}$, which are at distance equal to $\Delta_{2^N-\text{PSK}}$, and for which the value of $\bigoplus\limits_{s_j} \hat{Y}^i_{I(\mathcal{D}_j^i)}$ is not equal to that of $\bigoplus\limits_{s_k} \hat{Y}^i_{I(\mathcal{D}_j^i)}$ is denoted as $P(\mathcal{D}_j^i)$,
	\begin{align*}
		\label{signal_pair_count}
		P(\mathcal{D}_j^i) =  \big\{(s_j, s_k)  \ |\ & j < k, \ dist(s_j,s_k) = \Delta_{2^N-\text{PSK}} \text{ and } 
		\bigoplus\limits_{s_j}  \hat{Y}^i_{I(\mathcal{D}_j^i)} \neq \bigoplus\limits_{s_k} \hat{Y}^i_{I(\mathcal{D}_j^i)}\big\}
	\end{align*}
	
	\begin{thm}
		
		\label{thm:pe}
		Given an ICP and a chosen index code of length $N$, let the index-coded bits be  modulated using the $2^N$-PSK configuration $\mathcal{S}_{2^N,\mathcal{M}}$, and then transmitted over an AWGN channel. For this setting, if all the receivers have noiseless side information, the probability of error	$\mathcal{P}_e^{\mathcal{D}_j^i}$,  in decoding a requested message at any receiver using the decoding strategy $\mathcal{D}_j^i$ is upper bounded as :
		\begin{equation*}
			\mathcal{P}_e^{\mathcal{D}_j^i} < \frac{|P(\mathcal{D}_j^i)|}{2^N}2{Q}\left(\sqrt{2 N\left(\frac{E_b}{N_o}\right)\left(\sin^2\left(\frac{\pi}{2^N}\right)\right)}\right)
		\end{equation*}
		
		\begin{proof}
			At any receiver, the probability of error in estimating its requested message is due to the error in estimating the index-coded bits. This is because it is assumed that there is no error in the side information bits available at the receiver, and hence, they will not contribute to the error in the index decoding step. 
			
			For estimating the PSK symbol transmitted over an AWGN channel, minimum Euclidean distance decoding is performed at a receiver. At receiver $R_i$, when a transmitted symbol $s_j \in \mathcal{S}_{2^N,\mathcal{M}}$ is wrongly estimated as $s_k$, all index-coded bits need not be wrongly estimated. Now, we will derive an upper bound on the probability that $Y_{I(\mathcal{D}_j^i)}$ is decoded in error.

			We know that the probability that a symbol $s_m \in \mathcal{S}_{2^N,\mathcal{M}}$ transmitted over an AWGN channel is wrongly decoded as $s_{m^{'}} \in \mathcal{S}_{2^N,\mathcal{M}}$ is 
			$P_{Err(m \rightarrow m^{'})}=Q(d_{mm{'}}/ \sqrt{2N_o})$, where
			$d_{mm{'}} = dist(s_j,s_k)$, $N_o$ is the noise variance, and $Q(x)$ is the tail probability of a standard normal distribution. Considering all the $M =2^N$ symbols to be equally likely, the union bound for the average probability of error is obtained as
			\begin{equation}
				\label{unionbound}
				P_{Err_{avg}} < \frac{1}{M}\sum_{m=1}^{M} \sum_{\substack{\\ m{'} \neq m }}Q(d_{mm^{'}}/ \sqrt{2N_o}) 
			\end{equation}
			
			For a  $2^N$-PSK constellation, the minimum Euclidean distance is 
			$\Delta_{2^N-\text{PSK}}=2\sqrt{E_s \sin^2(\frac{\pi}{M})}$, where E$_s$ is the signal energy.
			In \eqref{unionbound}, the dominant terms will be those with $d_{mm^{'}}$ equal to the minimum Euclidean distance 	$\Delta_{2^N-\text{PSK}}$. In any $2^N$-PSK constellation, for a given transmitted symbol $s_m$, there will be two other symbols which are at a distance equal to $\Delta_{2^N-\text{PSK}}$ from $s_m$. Hence, we can approximate \eqref{unionbound} as 
			\[P_{Err_{avg}} < \frac{1}{M}\sum_{m=1}^{M} 2{Q}\left(\sqrt{2\left(\frac{E_s}{N_o}\right)\sin^2\left(\frac{\pi}{M}\right)}\right).\]
			Using the fact that the symbol energy $E_s$ is equal to $N$ times the bit energy $E_b$, the above equation can be re-written as 
			\begin{equation}
				\label{mpskerr}
				P_{Err_{avg}} < \frac{1}{M}\sum_{m=1}^{M} 2{Q}\left(\sqrt{2N\left(\frac{E_b}{N_o}\right)\sin^2\left(\frac{\pi}{M}\right)}\right).
			\end{equation}

			For determining the probability of error in decoding $W_i$, using a decoding strategy $\mathcal{D}_j^i$, at  $R_i$, we only need to consider the error events where the XOR of the bits used in decoding, i.e., $Y_{I(\mathcal{D}_j^i)}$, are in error. Up to first-order approximations, the probability that ${W}_i$ is estimated wrongly, $Pr(\hat{W}_i \neq W_i)$, is determined by the probability of the error events $(s_j \rightarrow s_k)$, i.e., $s_j$ decoded as $s_k$, for $(s_j,s_k) \in P(\mathcal{D}_j^i)$. Therefore, using \eqref{mpskerr} and the number of signal pairs in  $P(\mathcal{D}_j^i)$ corresponding to the error events of interest, the probability of error $\mathcal{P}_e^{\mathcal{D}_j^i}$ can be upper bounded as

			\begin{equation}
				\label{awgnperr}
				\mathcal{P}_e^{\mathcal{D}_j^i} < \frac{|P(\mathcal{D}_j^i)|}{2^N}2{Q}\left(\sqrt{2 N\left(\frac{E_b}{N_o}\right)\sin^2\left(\frac{\pi}{2^N}\right)}\right)
			\end{equation}
		\end{proof}
	\end{thm}

	\begin{cor}
		\label{cor:dc}
		For a given ICP and a chosen index code of length $N$, let $\mathcal{S}_{2^N,\mathcal{M}}$ be the PSK signal set configuration used for transmitting the index code over an AWGN channel. Under this setting, at a receiver $R_i$, an optimal decoding strategy w.r.t probability of error performance is given by $\mathcal{D}^i_* = \arg \min\limits_{ j \in [r]}(|P(\mathcal{D}^i_j)|)$, where $r$ is the number of decoding strategies at $R_i$.
		
		\begin{proof}
			From Theorem \ref{thm:pe}, up to first-order approximations, the probability of error in decoding $W_i$ using a decoding strategy $\mathcal{D}_j^i$  at $R_i$ is determined by the error events $(s_j \rightarrow s_k)$, i.e., $s_j$ decoded as $s_k$, for $(s_j,s_k) \in P(\mathcal{D}_j^i)$.  Hence, an optimal decoding strategy is  one that minimizes the number of signal pairs in $P(\mathcal{D}_j^i)$. 
		\end{proof}
	\end{cor}

	\begin{cor}
		\label{cor:Num_bits}
		For a given ICP and a chosen index code of length $N$ with receivers having noiseless side information, when the index-coded bits are transmitted employing any $2^N$-PSK configuration, the probability of  error in decoding a requested message at any receiver does not depend on the number of index-coded bits used to decode it.
		\begin{proof}
			From \eqref{awgnperr}, we saw that the probability of error for a decoding strategy $\mathcal{D}_j^i$ depends on $|P(\mathcal{D}_j^i)|$ and not on the number of estimated bits in the linear combination used in $\mathcal{D}_j^i$. Consider the $8$- PSK constellation with decimal to binary mapping shown in Fig. \ref{8-psk}. It can be seen that when $\hat{Y}^i_{I(\mathcal{D}_j^i)} = \{\hat{y}^i_1\}$, the number of error events $(s_j \rightarrow s_k)$, with  $(s_j,s_k) \in P(\mathcal{D}_j^i)$ is two, whereas with $\hat{Y}^i_{I(\mathcal{D}_j^i)} = \{\hat{y}^i_3\}$, the number of such error events  is $8$. Hence, a decoding strategy using $\hat{y}^i_1$ will not perform the same as another one using $\hat{y}^i_3$. Further, consider that $\mathcal{D}_j^i$ uses a linear combination of  $\hat{y}^i_1$ and $\hat{y}^i_2$, for which the number of error events $\{(s_j \rightarrow s_k), (s_j,s_k) \in  P(\mathcal{D}_j^i)\}$ is two which implies that this strategy will perform better than the one using $\hat{y}^i_3$ alone. Hence, we see that the probability of error performance is not dependent on the number of index-coded bits used in decoding. 
		\end{proof} 
	\end{cor}

	The above result is true for any multi-level modulation scheme. Based on the selection criterion in the Corollary \ref{cor:dc} above, we now propose the following algorithm for finding an optimal decoding strategy $\mathcal{D}^i_*$ at receiver $R_i$. 
	
	\begin{algorithm}
		\caption{Find the best decoding strategy at $R_i$}
		
		\begin{algorithmic}[1]
			
			\Require Encoding matrix $\mathbf{L}$, side information $K_i$ and chosen PSK configuration $\mathcal{S}_{2^N,\mathcal{M}}$. 
			\Ensure Best Decoding strategy, $\mathcal{D}^i_*$
			\State Find all decoding strategies at $R_i$, say $\mathcal{D}^i_1, \mathcal{D}^i_2, \cdots, \mathcal{D}^i_r$.
			\If {r==1}
			\State \textbf{return} $\mathcal{D}^i_* = \mathcal{D}^i_1$. 
			\EndIf
			\ 
			\State Determine $P(\mathcal{D}^i_1), P(\mathcal{D}^i_2), \cdots P(\mathcal{D}^i_r)$.
			
			\State Compute $\mathbf{\mathcal{D}}_{min}^i =\arg \min\limits_{ j \in [r]}(|P(\mathcal{D}^i_j)|)$.
			
			\State  $\mathcal{D}^i_* = \mathcal{D}^i_j$, for some  $\mathcal{D}^i_j \in \mathbf{\mathcal{D}}_{min}^i $. 
			
		\end{algorithmic}
		\label{euclid}
	\end{algorithm}
	
	\begin{rem}
		\label{remark}
		For a receiver $R_i$, if there is more than one decoding strategy in the set  $\mathbf{\mathcal{D}}_{min}^i$ computed in step 6 of Algorithm \ref{euclid}, any one of them can be chosen arbitrarily for obtaining the best probability of error performance. 
	\end{rem}
	
		\subsection{Illustrative Example}
	\label{sec:example1}
	Consider the ICP and the chosen index code of length $N=3$ in Example \ref{example-1}, for which the chosen $8$-PSK configuration is given in Fig. \ref{8-psk}. The minimum Euclidean distance of the unit-energy $8$-PSK constellation is $\Delta_{8-\text{PSK}}=0.7653$. Now, consider the performances of various decoding strategies at different receivers. 
	
\begin{figure}\centering

		\captionsetup{justification = centering}
		\includegraphics[width = 0.65\columnwidth]{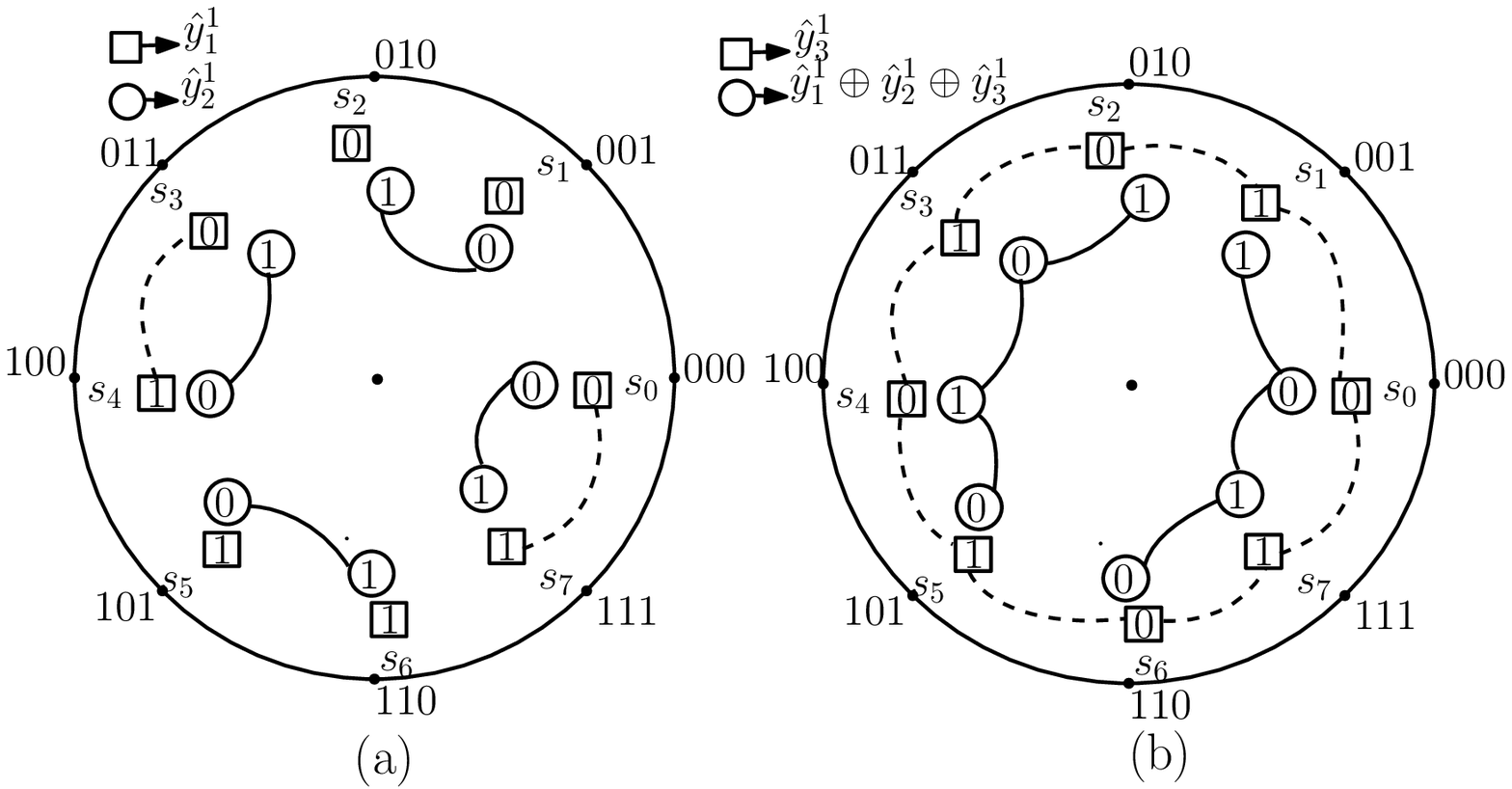}
		\caption{Decoding at $R_1$: (a) $P(\mathcal{D}^1_1)$, $P(\mathcal{D}^1_2)$ (b) $P(\mathcal{D}^1_3)$,  $P(\mathcal{D}^1_4)$}
		\label{rx1}
		
	\end{figure}

	At $R_1$, there are a total of four decoding strategies, for each of which the set $P(\mathcal{D}^1_j), \ j \in [4]$ are shown in Fig. \ref{rx1}. We can see that, for $\mathcal{D}^1_1$, there are two signal pairs at minimum Euclidean distance differing in the value of $\hat{y}_1^1$, i.e., $P(\mathcal{D}^1_1) = \{(s_0,s_7), (s_3,s_4)\}$. Similarly, there are four signal pairs at minimum Euclidean distance differing in the value of $\hat{y}_2^1$, which implies that $|P(\mathcal{D}^1_2)| = 4$. For the decoding strategy $\mathcal{D}^1_3$ using the index-coded bit $\hat{y}_3^1$ for decoding $x_1$, $|P(\mathcal{D}^1_3)| = 8$, whereas for $\mathcal{D}^1_4$ using the linear combination $\hat{y}^1_1\oplus\hat{y}^1_2\oplus\hat{y}^1_3$ of the index-coded bits along with the side information $K_1$, $|P(\mathcal{D}^1_4)| = 6$. Hence, using Corollary \ref{cor:dc}, we find that the optimal decoding strategy at receiver $R_1$ is $\mathcal{D}^1_1$.

\begin{figure}\centering

		\captionsetup{justification = centering}
		\includegraphics[width=0.65\columnwidth]{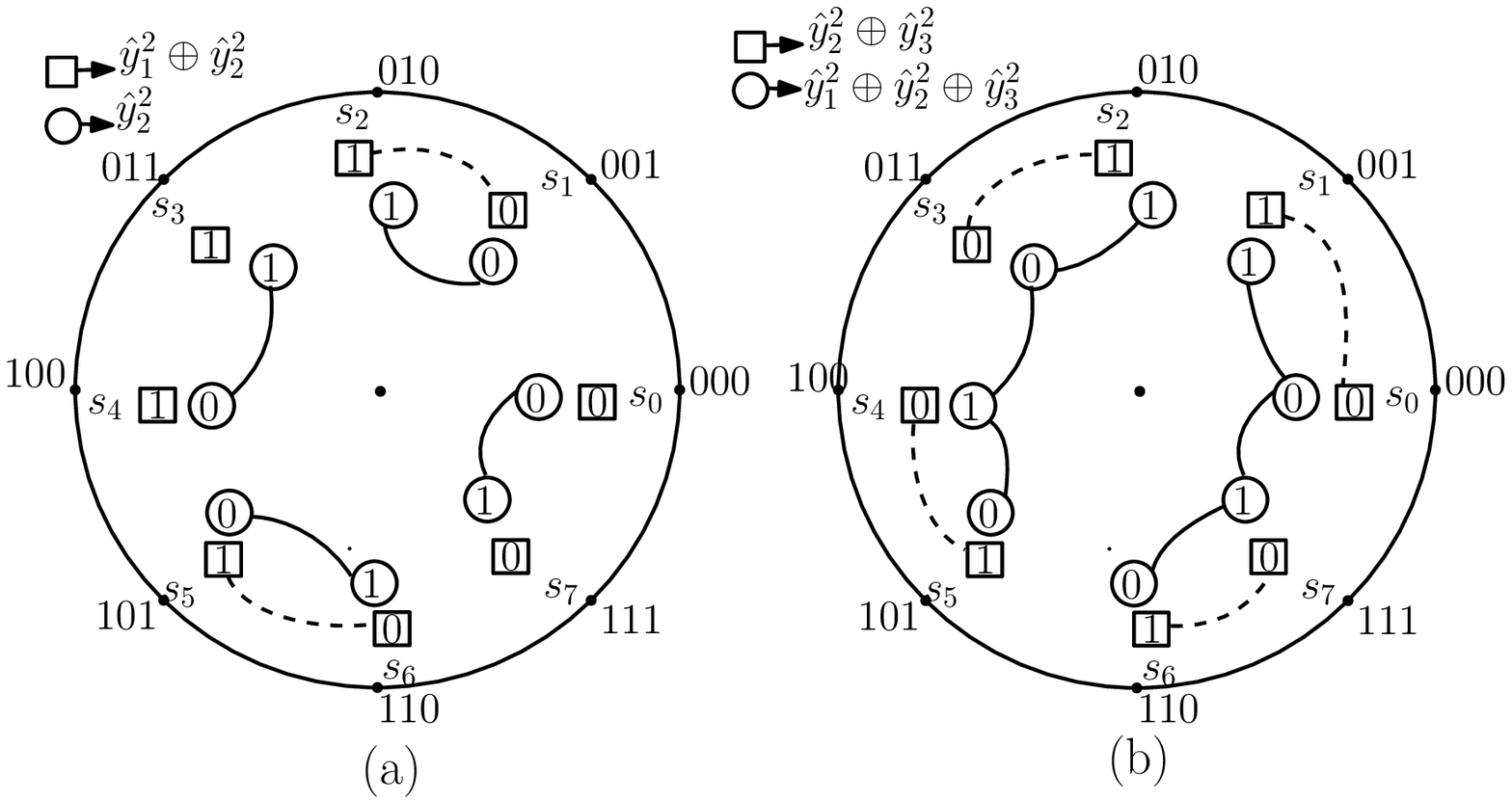}
		\caption{Decoding at $R_2$: (a) $P(\mathcal{D}^2_1)$, $P(\mathcal{D}^2_2)$ (b) $P(\mathcal{D}^2_3)$,  $P(\mathcal{D}^2_4)$}
		\label{rx2}
	\end{figure}

	Similarly, at $R_2$, as shown in Fig. \ref{rx2}, the number of signal pairs at minimum Euclidean distance differing in the value of $\hat{y}^2_1 \oplus \hat{y}^2_2$ used by $\mathcal{D}^2_1$ is two, whereas, for the strategy  $\mathcal{D}^2_2$, it has four signal pairs in $P(\mathcal{D}^2_2)$ given by $\{(s_0,s_7), (s_1,s_2),(s_3,s_4),(s_5,s_6)\}$. Likewise, for $\mathcal{D}^2_3$ using the linear combination $\hat{y}^2_2\oplus\hat{y}^2_3$ of estimated index-coded bits, $|P(\mathcal{D}^2_3)| = 4$, whereas for $\mathcal{D}^2_4$ using $\hat{y}^2_1\oplus\hat{y}^2_2 \oplus \hat{y}^2_3$, we have $|P(\mathcal{D}^2_4)| = 6$. Hence, the optimal decoding strategy is $\mathcal{D}^2_1$.

	\begin{figure*}\centering		
		
		\captionsetup{justification = centering}
		\includegraphics[width=1\columnwidth]{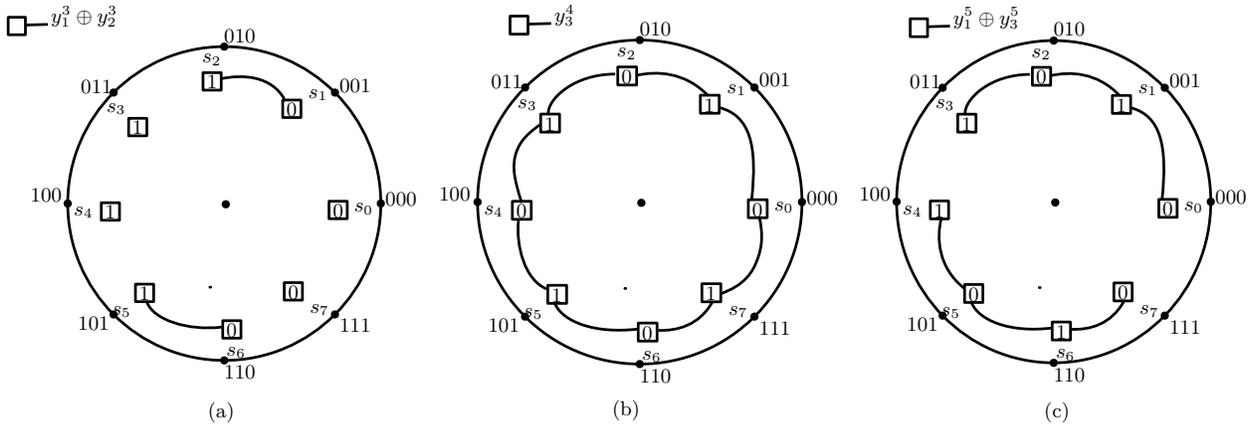}
		\caption{(a) Decoding at $R_3$: $P(\mathcal{D}^3_1)$ (b) Decoding of $R_4$: $P(\mathcal{D}^4_1)$ and (c) Decoding at $R_5$: $P(\mathcal{D}^5_1)$.}
		\label{rx3_4_5}
		
	\end{figure*}
	
	There is only one decoding strategy each at receivers $R_3$, $R_4$, and $R_5$ for decoding their requested messages. For the $8$-PSK configuration shown in Fig. \ref{8-psk}, for each of these decoding strategies, the set of signal pairs at the minimum Euclidean distance differing in the value of XOR of the index-coded bits used is given in Fig. \ref{rx3_4_5}. From this figure, we can see that $|P(\mathcal{D}^3_1)| = 2$, $|P(\mathcal{D}^4_1)| = 8$ and $|P(\mathcal{D}^5_1)| = 6$.
	\section{Receivers with Noisy Side Information}
	\label{noisy_si}
	\subsection{System Model}
	\label{model2}  
	We will consider the same system model as discussed in Section \ref{model1} which has a central server with access to a library $\mathcal{X}=\{x_1, x_2,\ldots, x_m\}$, $x_i\in\ \mathbb{F}_2$, of messages, broadcasts to a set of $n$ receivers, $\mathcal{R}=\{R_1, R_2,\ldots, R_n\}$. Here also, we will use linear index codes. Since we are considering that the side information at the receivers are noisy, we assume that the transmissions take place in two phases. In the first phase, which happens during a window with minimal network congestion, the server broadcasts each message in $\mathcal{X}$ independently after BPSK modulation over a noisy channel. The noise characteristics of the channel between the source and each of the receivers are assumed to be identical. Hence, the signal-to-noise ratio (SNR) observed at each receiver is the same and is denoted as $\Gamma_{si}$. Depending on the received signal levels, some symbols are decoded at a receiver while others are erased.  Hence, at the end of this phase of message broadcast, a receiver $R_i$ will have a set $K_i$ of decoded messages as side information.

	Assuming that the index code transmissions happen when the network is heavily congested, for further reduction in bandwidth, the length-$N$ index-coded vector $\mathbf{y} = (y_1, y_2, \ldots, y_N)$ is transmitted after $2^N$-PSK modulation. For an AWGN broadcast channel, at $R_i$, the received complex symbol is $c_i = \mathcal{M}(\mathbf{y}) +n_i$, where $\mathcal{M}(\cdot)$ denotes the mapping of index-coded bits to a PSK constellation, and $n_i$ is the white Gaussian noise. The signal-to-noise ratio at which this symbol $c_i$ is received is denoted as $\Gamma_{ic}$ which is equal to $\frac{E_s}{N_o}$ or $\frac{NE_b}{N_o}$ with $E_s$ being the symbol energy, $E_b$ the energy per bit and $N_o$ being the variance of the additive noise $n_i$. In either case, after minimum distance decoding at a receiver $R_i$, the vector of estimated index-coded bits is denoted as  $\hat{\mathbf{y}}^i = (\hat{y}^i_1, \hat{y}^i_2, \ldots, \hat{y}^i_N)$. 
	
	Similar to noiseless side information case, for a linear encoding scheme, the there will be many decoding function $\mathcal{D}^i$ at receiver $R_i$ is a linear combination of all or a subset of the index-coded bits and a subset of its side information. We will now discuss an example.

	\begin{exmp}
		\label{example-2}
		Consider an ICP with  $m=n=5$ and $W_i=x_i \ \forall{i} \in \{1,2,3,4,5\}$. $\mathcal{X}$ is transmitted using BPSK over an AWGN channel in low-traffic hours. Depending upon the threshold value or sensitivities at each receiver, it decodes some of the messages while other messages get erased, which results in the following side information at the receivers $K_1=\{\tilde{x}_2,\tilde{x}_3,\tilde{x}_4,\tilde{x}_5\}$, $K_2=\{\tilde{x}_1,\tilde{x}_3,\tilde{x}_4,\tilde{x}_5\}$, $K_3=\{\tilde{x}_2, \tilde{x}_4\}$, $K_4=\{\tilde{x}_1, \tilde{x}_3\}$, and $K_5=\{\tilde{x}_3\}$. For this ICP, the length of an optimal linear index code is $N=3$.  We consider the linear encoding scheme specified by 
		$\mathbf{L} = \begin{bmatrix}
			1 & 0  & 0 & 1 & 1 \\
			1 & 1 & 1 & 1 & 1 \\
			1 & 0 & 1 & 1 & 0
		\end{bmatrix}^T$ and the index coded vector is $\mathbf{y}=(y_1,y_2,y_3)$ where $\mathbf{y}=\mathbf{x}\mathbf{L}$, for $\mathbf{x} \in \mathbb{F}_2^5$. We transmit the above index code using the 8-PSK mapping given in Fig.\ref{8-psk} through an AWGN channel. At the receiver $R_i$, after performing ML decoding on the received signal, the vector of estimated index coded bits is $\mathbf{\hat{y}}^i= (\hat{y}^i_1, \hat{y}^i_2,\hat{y}^i_3)$. The different decoding strategies at each of the receivers are given below. 
		
		At $R_1$:
		\begin{itemize}
			\item[]$\hat{x}_1=\mathcal{D}_1^1(\mathbf{\hat{y}}^1,K_1) := \hat{y}^1_1  \oplus \tilde{x}_4  \oplus \tilde{x}_5$
			\item[]$\hat{x}_1=\mathcal{D}_2^1(\mathbf{\hat{y}}^1,K_1) :=\hat{y}^1_2 \oplus \tilde{x}_2 \oplus \tilde{x}_3 \oplus \tilde{x}_4 \oplus \tilde{x}_5$
			\item[]$\hat{x}_1=\mathcal{D}_3^1(\mathbf{\hat{y}}^1,K_1) :=\hat{y}^1_3 \oplus \tilde{x}_3 \oplus \tilde{x}_4$
			\item[]$\hat{x}_1=\mathcal{D}_4^1(\mathbf{\hat{y}}^1,K_1) :=\hat{y}^1_1 \oplus \hat{y}^1_2 \oplus \hat{y}^1_3 \oplus \tilde{x}_4 \oplus \tilde{x}_2$
		\end{itemize}
		
		At $R_2$:
		\begin{itemize}
			\item[]$\hat{x}_2=\mathcal{D}_1^2(\mathbf{\hat{y}}^2,K_2) :=\hat{y}^2_1 \oplus \hat{y}^2_2 \oplus \tilde{x}_3$
			\item[]$\hat{x}_2=\mathcal{D}_2^2(\mathbf{\hat{y}}^2,K_2) :=\hat{y}^2_2 \oplus \tilde{x}_1 \oplus \tilde{x}_3 \oplus \tilde{x}_4 \oplus \tilde{x}_5$
			\item[]$\hat{x}_2=\mathcal{D}_3^2(\mathbf{\hat{y}}^2,K_2) :=\hat{y}^2_2 \oplus \hat{y}^2_3 \oplus \tilde{x}_5$
			\item[]$\hat{x}_2=\mathcal{D}_4^2(\mathbf{\hat{y}}^2,K_2) :=\hat{y}^2_1 \oplus \hat{y}^2_2 \oplus \hat{y}^2_3 \oplus \tilde{x}_1 \oplus \tilde{x}_4$
		\end{itemize}
		
		At $R_3$:
		\begin{itemize}
			\item[]$\hat{x}_3=\mathcal{D}_1^3(\mathbf{\hat{y}}^3,K_3) :=\hat{y}^3_1 \oplus \hat{y}^3_2 \oplus \tilde{x}_2$
		\end{itemize}
		
		At $R_4$:
		\begin{itemize}
			\item[]$\hat{x}_4=\mathcal{D}_1^4(\mathbf{\hat{y}}^4,K_4) :=\hat{y}^4_3 \oplus \tilde{x}_1 \oplus \tilde{x}_3$
		\end{itemize}
		
		At $R_5$:
		\begin{itemize}
			\item[]$\hat{x}_5=\mathcal{D}_1^5(\mathbf{\hat{y}}^5,K_5) :=\hat{y}^5_1 \oplus \hat{y}^5_3 \oplus \tilde{x}_3$.
		\end{itemize}
	\end{exmp}

	From the above example, we saw that for a given index code, there could exist different ways in which a receiver could decode its requested message. We will derive a criterion for choosing an optimal decoding strategy at a receiver. 
	
	\subsection{Main Results} 
	\label{mainresult2}
	
	For a chosen index code of length $N$ of a given ICP and a chosen mapping $\mathcal{M}$ of index codewords to PSK signal points, let the $2^N$- PSK signal set configuration used for transmission be denoted as $\mathcal{S}_{2^N,\mathcal{M}}$. At receiver $R_i$ which has the side information  $K_i$ , denote the set $\{\hat{y}^i_j\}_{j \in [N]}$ of estimated index-coded bits by $\hat{Y}^i$. For a decoding strategy $\mathcal{D}_j^i$ at $R_i$, let the indices of the index-coded bits used by $\mathcal{D}_j^i$ be denoted as $I(\mathcal{D}_j^i)$. Similarly, the index set of the subset of the side information $K_i$ used by $\mathcal{D}_j^i$ is represented as $S(\mathcal{D}_j^i)$. Hence, from the set $\hat{Y}^i$ of estimated index coded bits, $\mathcal{D}_j^i$ uses a linear combination of the subset $\hat{Y}^i_{I(\mathcal{D}_j^i)} \subseteq \hat{Y}^i$  of estimated index-coded bits, along with the subset $\tilde{\mathcal{X}}_{S(\mathcal{D}_j^i)} \subseteq K_i$ of the side information to decode $W_i$. 
	The XOR of the bits in $\hat{Y}^i_{I(\mathcal{D}_j^i)}$ corresponding to a signal point $s_k \in \mathcal{S}_{2^N,\mathcal{M}}$ is denoted as $\bigoplus\limits_{s_k} \hat{Y}^i_{I(\mathcal{D}_j^i)}$.

	\begin{thm}
		\label{thm:noisyPe}
		For a given ICP, let the side information at the receivers be noisy. For a chosen index code of length $N$, let $\mathcal{S}_{2^N,\mathcal{M}}$ be the PSK configuration used for transmitting the index code over an AWGN channel. In this setting, the probability of bit error obtained using a decoding strategy $\mathcal{D}_j^i$ is given by
		\begin{equation}
			\label{eq:Thm_NosiyPe}
			\mathcal{P}_e^{\mathcal{D}_j^i}=\mathcal{P}_e^{\bigoplus \hat{Y}^i_{I(\mathcal{D}_j^i)}}(1-2\mathcal{P}_e^{si})^{|S(\mathcal{D}^i_j)|}+\frac{1-(1-2\mathcal{P}_e^{si})^{|S(\mathcal{D}^i_j)|}}{2},
		\end{equation}
		where $\mathcal{P}_e^{si}$ is the probability of bit error for BPSK-modulated message transmissions over an AWGN channel, and $\mathcal{P}_e^{\bigoplus \hat{Y}^i_{I(\mathcal{D}_j^i)}}$ is the probability of error for the linear combination $\bigoplus \hat{Y}^i_{I(\mathcal{D}_j^i)}$ of the index-coded bits obtained from the $M$-PSK signal.	
		\begin{proof}		
			Any decoding strategy, in a general form, can be written as
			$\mathcal{D}_j^i=\bigoplus \hat{Y}^i_{I(\mathcal{D}_j^i)} \bigoplus \tilde{\mathcal{X}}_{S(\mathcal{D}_j^i)}$.
			At a receiver, decoding the $M$-PSK symbol corresponding to the index-coded bits and the side information messages take place independent of each other. Since the error analysis is done over the binary field  $\mathbb{F}_2$, a decoding strategy $\mathcal{D}_j^i$ is said to make an error in estimating the demanded message when there are an odd number of errors in the decoded side information bits and the linear combination $\bigoplus \hat{Y}^i_{I(\mathcal{D}_j^i)}$ used by $\mathcal{D}_j^i$. The side information bits are transmitted using BPSK modulation, so they are independently of each other. Hence, $\mathcal{D}_j^i$ will decode wrongly in the following two situations:
			\begin{itemize}
				\item The linear combination $\bigoplus \hat{Y}^i_{I(\mathcal{D}_j^i)}$ of index-coded bits is decoded wrongly, and an even number of errors in $S(\mathcal{D}_j^i)$.
				\item The linear combination $\bigoplus \hat{Y}^i_{I(\mathcal{D}_j^i)}$ of index-coded bits is decoded correctly, and an odd number of errors in $S(\mathcal{D}_j^i)$.
			\end{itemize} 	
			\begin{multline}
				\label{perr}
				\mathcal{P}_e^{\mathcal{D}_j^i}=\mathcal{P}_e^{\bigoplus \hat{Y}^i_{I(\mathcal{D}_j^i)}}\Bigg(\sum_{\substack{v \le |S(\mathcal{D}_j^i)|  \\ v \ \text{even}}} \binom{|S(\mathcal{D}_j^i)|}{v} (\mathcal{P}_e^{si})^v  (1-\mathcal{P}_e^{si})^{|S(\mathcal{D}_j^i)|-v}\Bigg)+(1-\mathcal{P}_e^{\bigoplus \hat{Y}^i_{I(\mathcal{D}_j^i)}})\\
				\Bigg(\sum_{\substack{v \le |S(\mathcal{D}_j^i)| \\ v \ \text{odd}}} \binom{|S(\mathcal{D}_j^i)|}{v}  (\mathcal{P}_e^{si})^v(1-\mathcal{P}_e^{si})^{|S(\mathcal{D}_j^i)|-v}\Bigg) 
			\end{multline} 			
			Using the binomial expansions for $(1+x)^{\eta}$ and $(1-x)^{\eta}$ with $x=\frac{\mathcal{P}_e^{si}}{1-\mathcal{P}_e^{si}}$ and $\eta = |S(\mathcal{D}_j^i)|$, we can simplify the above expression as follows.	
			\begin{equation*}
				\mathcal{P}_e^{\mathcal{D}_j^i}=\mathcal{P}_e^{\bigoplus \hat{Y}^i_{I(\mathcal{D}_j^i)}}(1-2\mathcal{P}_e^{si})^\eta+\frac{1-(1-2\mathcal{P}_e^{si})^\eta}{2}
			\end{equation*}
		where,
		\begin{equation*}
			\label{side_info_perr}
			\mathcal{P}_e^{si} < {Q}(\sqrt{2\Gamma_{si}})
		\end{equation*}
				\begin{equation*}
				\label{perr_awgn}
				\mathcal{P}_e^{\bigoplus \hat{Y}^i_{I(\mathcal{D}_j^i)}} <\frac{|P(\mathcal{D}_j^i)|}{2^N}2{Q}\Bigg(\sqrt{2 {\Gamma}_{ic}\Big(\sin^2\Big(\frac{\pi}{2^N}\Big)\Big)}\Bigg)
			\end{equation*}	
		\end{proof}
	\end{thm}
	
	We can see that the expression for the probability of bit error in \eqref{eq:Thm_NosiyPe} derived in Theorem \ref{thm:noisyPe} consists of two terms, one determined by the error in the index-coded $M$-PSK transmission and the other by the error in the BPSK-modulated message transmissions from which the side information at a receiver is obtained. In the following lemma, we determine the value of $\Gamma_{ic}$ at which the first term in \eqref{eq:Thm_NosiyPe} becomes equal to the second term in \eqref{eq:Thm_NosiyPe}. We denote this value of $\Gamma_{ic}$, which we call the threshold value, as $\Gamma^{th}$. The significance of $\Gamma^{th}$ lies in the fact that increasing $\Gamma_{ic}$ beyond $\Gamma^{th}$ does not improve the probability of error performance any further. 
	\begin{lem}
		\label{lemma_ebnoth}
		For a given ICP and a chosen index code of length $N$, consider that the index-coded bits are transmitted using $2^N$-PSK modulation over an AWGN channel. Let the messages in the side information of a receiver $R_i$ be obtained with an SNR of $\Gamma_{si}$ through BPSK-modulated transmissions over an AWGN channel. In this scenario, the threshold value of $\Gamma_{ic}$ for a given decoding strategy $\mathcal{D}_j^i$ is given by
		\begin{equation}
			\label{ebnoth}
			\Gamma^{th}(\mathcal{D}_j^i) \approx \beta\Bigg({Q}^{-1}\Bigg(\frac{2^N(1-(1-2\mathcal{P}_e^{si})^{|S(\mathcal{D}^i_j)|})}{P(\mathcal{D}_j^i)2(1-2\mathcal{P}_e^{si})^{|S(\mathcal{D}^i_j)|}}\Bigg)\Bigg)^2,
		\end{equation}
		where $\beta=1/\Big(2N \Big(\sin^2(\frac{\pi}{2^N})\Big)\Big)$
		
		\begin{proof}
			At a given receiver $R_i$, the probability of bit error performance of a decoding strategy $\mathcal{D}_j^i$ is given by \eqref{eq:Thm_NosiyPe}. The second term in \eqref{eq:Thm_NosiyPe} is independent of $\Gamma_{ic}$. At very low values of $\Gamma_{ic}$ compared to $\Gamma_{si}$, the error performance of $\mathcal{D}_j^i$ is determined predominantly by the first term. As the value of $\Gamma_{ic}$ increases, the value of the first term reduces exponentially for transmissions over an AWGN channel. As long as the order of the first term is greater than that of the second term, the probability of bit error will continue to trace the waterfall curve with an increasing value of $\Gamma_{ic}$. Whereas, when the value of the first term is approximately equal to that of the second term, the probability of error curve does not fall significantly with further increase in the value of $\Gamma_{ic}$. An approximate value of $\Gamma_{ic}$ at which this ``error floor'' occurs can be estimated using \eqref{eq:Thm_NosiyPe} as the value at which
			\[ \mathcal{P}_e^{\bigoplus \hat{Y}^i_{I(\mathcal{D}_j^i)}}(1-2\mathcal{P}_e^{si})^{|S(\mathcal{D}^i_j)|} \approx \frac{1-(1-2\mathcal{P}_e^{si})^{|S(\mathcal{D}^i_j)|}}{2}, \]
			which can simplified using \eqref{perr_awgn} for transmissions over AWGN channels to get 
			\begin{equation*}
				\Gamma^{th}(\mathcal{D}_j^i) \approx \beta\Bigg({Q}^{-1}\Bigg(\frac{2^N(1-(1-2\mathcal{P}_e^{si})^{|S(\mathcal{D}^i_j)|})}{P(\mathcal{D}_j^i)2(1-2\mathcal{P}_e^{si})^{|S(\mathcal{D}^i_j)|}}\Bigg)\Bigg)^2
			\end{equation*}
			where $\beta=\frac{1}{2N \Big(\sin^2(\frac{\pi}{2^N})\Big)}$, and $\mathcal{P}_e^{si}={Q}(\sqrt{2\Gamma_{si}})$.
			From \eqref{ebnoth}, it can be seen clearly that as the value of $\Gamma_{si}$ increases, the value of $\Gamma^{th}(\mathcal{D}_j^i)$ also  increases.
		\end{proof}
	\end{lem}
	
	\begin{thm}
		\label{thm:OptDS}
		For the setting considered in Theorem \ref{thm:noisyPe}, at a receiver $R_i$ which has $r$ possible decoding strategies, an optimal decoding strategy w.r.t probability of error is given by $$\mathcal{D}^i_* = \begin{cases}
			\arg \min\limits_{ j \in [r]}(|S(\mathcal{D}^i_j)|),\Gamma_{ic} \ge \max \{\Gamma^{th}(\mathcal{D}_1^i), \cdots, \Gamma^{th}(\mathcal{D}_r^i)\} \\
			\arg \min\limits_{ j \in [r]}(|P(\mathcal{D}^i_j)|),\text{ otherwise}.
		\end{cases}$$	    
		\begin{proof}
			\textbf{Case 1: Transmission of side information is done in the low $\Gamma_{si}$ regime}\\
			When the value of $\Gamma_{si}$ is considerably lower than that of $\Gamma_{ic}$, the order of the second term is higher than that of the first term in \eqref{eq:Thm_NosiyPe}. Therefore, the error performance is primarily determined by the second term, and we can approximate the probability of bit error obtained by $\mathcal{D}^i_j$ as
			\begin{equation}
				\label{lowsnr}
				\mathcal{P}_e^{\mathcal{D}_j^i} \approx \frac{1-(1-2\mathcal{P}_e^{si})^{|S(\mathcal{D}^i_j)|}}{2}
			\end{equation} 
			
			In this regime, the probability of error will not decrease significantly with an increase in $\Gamma_{ic}$, which can be inferred from the approximate expression in \eqref{lowsnr}. 
			Thus, at low $\Gamma_{si}$ an optimal decoding strategy at $R_i$ will be given by $\mathcal{D}^i_* = \arg \min\limits_{ j \in [r]}(|S(\mathcal{D}^i_j)|)$.
			
			\textbf{Case 2: Transmission of side information is done in the high $\Gamma_{si}$ regime.}
			\\
			In this regime, in \eqref{eq:Thm_NosiyPe}, the order of the second term is lower than that of the first term. Therefore, the expression for the probability of bit error can be approximated as
			\begin{equation}
				\label{eq:highsnr}
				\mathcal{P}_e^{\mathcal{D}_j^i} \approx \mathcal{P}_e^{\bigoplus \hat{Y}^i_{I(\mathcal{D}_j^i)}}(1-2\mathcal{P}_e^{si})^{|S(\mathcal{D}^i_j)|}.
			\end{equation}
			Hence, in this regime, where the probability that a message in the side information of $R_i$ is decoded in error is very low, an optimal decoding strategy will be the one that minimizes the number of signal pairs in $P(\mathcal{D}^i_j)$, i.e., 
			$\mathcal{D}^i_* = \arg \min\limits_{ j \in [r]}(|P(\mathcal{D}^i_j)|)$
		\end{proof}
	\end{thm}

	Based on the selection criterion in the Theorem \ref{thm:OptDS} above, we now propose the algorithm for finding an optimal decoding strategy $\mathcal{D}^i_*$ at receiver $R_i$ in Algorithm.\ref{noisy_algo} 
	
	\begin{algorithm}
		\caption{Find the best decoding strategy at $R_i$}
		
		\begin{algorithmic}[1]
			
			\Require Encoding matrix $\mathbf{L}$, side information $K_i$, the chosen PSK configuration $\mathcal{S}_{2^N,\mathcal{M}}$, and the side information SNR $\Gamma_{si}$. 
			\Ensure Best Decoding strategy, $\mathcal{D}^i_*$
			\State Find all decoding strategies at $R_i$, say $\mathcal{D}^i_1, \mathcal{D}^i_2, \cdots, \mathcal{D}^i_r$.
			\State Calculate the threshold values ${\Gamma^{th}(\mathcal{D}_1^i), \Gamma^{th}(\mathcal{D}_2^i), \cdots, \Gamma^{th}(\mathcal{D}_r^i)}$.
			\If {r==1}
			\State \textbf{return} $\mathcal{D}^i_* = \mathcal{D}^i_1$. 
			\EndIf
			\ 
			\If {$\Gamma_{ic}$ $\ge$ $\max \{\Gamma^{th}(\mathcal{D}_1^i), \Gamma^{th}(\mathcal{D}_2^i) \cdots \Gamma^{th}(\mathcal{D}_r^i)\}$}
			\State Determine $|S(\mathcal{D}^i_1)|, |S(\mathcal{D}^i_2)|, \cdots |S(\mathcal{D}^i_r)|$.
			\State Compute $\mathbf{\mathcal{D}}_{min}^i =\arg \min\limits_{ j \in [r]}(|S(\mathcal{D}^i_j)|)$.
			\If {$|\mathcal{D}_{min}^i|=z \gneq 1$}
			\State Determine $P(\mathcal{D}^i_{min}(1)), \cdots, P(\mathcal{D}^i_{min}(z))$.
			
			\State Compute $\mathbf{\mathcal{D}}_{min}^i =\arg \min\limits_{ j \in [z]}(|P(\mathcal{D}^i_{min}(j))|)$.

			\EndIf
			\State  $\mathcal{D}^i_* = \mathcal{D}^i_j$, for some  $\mathcal{D}^i_j \in \mathbf{\mathcal{D}}_{min}^i $. 
			\Else 
			\State Determine $P(\mathcal{D}^i_1), P(\mathcal{D}^i_2), \cdots P(\mathcal{D}^i_r)$.
			
			\State Compute $\mathbf{\mathcal{D}}_{min}^i =\arg \min\limits_{ j \in [r]}(|P(\mathcal{D}^i_j)|)$.
			\If {$|\mathcal{D}_{min}^i|=z \gneq 1$}
			\State Determine $S(\mathcal{D}^i_{min}(1)), \cdots, S(\mathcal{D}^i_{min}(z))$.
			
			\State Compute $\mathbf{\mathcal{D}}_{min}^i =\arg \min\limits_{ j \in [z]}(|S(\mathcal{D}^i_{min}(j))|)$. 
			\EndIf

			\State  $\mathcal{D}^i_* = \mathcal{D}^i_j$, for some  $\mathcal{D}^i_j \in \mathbf{\mathcal{D}}_{min}^i $. 
			\EndIf

		\end{algorithmic}
		\label{noisy_algo}
	\end{algorithm}
	
	\begin{rem}
		\label{remark_algo2}
		For a receiver $R_i$, if there is more than one decoding strategy in the set  $\mathbf{\mathcal{D}}_{min}^i$ computed in step 13 or step 21 of Algorithm \ref{noisy_algo}, any one of them can be chosen arbitrarily for obtaining the best probability of error performance. 
		
	\end{rem}
	
		\subsection{Illustrative Example}
	\label{sec2:example}
	
	We will analyze the ICP described in Example \ref{example-1}, for which the chosen $8$-PSK configuration is given in Fig. \ref{8-psk}. Let us assume that the transmission of side information is done in a low SNR regime, say, at $\Gamma_{si} = 5$ dB. At $R_1$, there are a total of $4$ decoding strategies. According to Theorem \ref{thm:OptDS}, an optimal decoding strategy will be one that uses the least number of side information bits. For $R_1$, $|S(\mathcal{D}_1^1)|= 2$, $|S(\mathcal{D}_2^1)|=4$, $|S(\mathcal{D}_3^1)|= 2$ and $|S(\mathcal{D}_4^1)| = 2$. Each of the decoding strategies $\mathcal{D}_1^1$, $\mathcal{D}_3^1$, and $\mathcal{D}_4^1$ uses the same number of side information bits which is the least among all possible strategies at $R_1$. According to Algorithm \ref{noisy_algo}, $\mathcal{D}_1^1$ will give the best performance as $|P(\mathcal{D}_1^1)|=2$ which is the minimum among the three strategies mentioned above. At $R_2$, there are a total of $4$ decoding strategies, $|S(\mathcal{D}_1^2)|=1$, $|S(\mathcal{D}_2^2)|=4$, $|S(\mathcal{D}_3^2)|=1$, and $|S(\mathcal{D}_4^2)|=2$. Since $|S(\mathcal{D}_1^2)|= |S(\mathcal{D}_3^2)|=1$, $\mathcal{D}_1^2$ will be returned as the optimal decoding strategy by Algorithm \ref{noisy_algo} as  we have $|P(\mathcal{D}_1^2)|<  |P(\mathcal{D}_3^2)|$. There is only one decoding strategy each at the receivers $R_3$, $R_4$, and $R_5$ for decoding their respective requested messages. While the receivers $R_3$ and $R_5$ use only one side information bit each, the receiver $R_4$ uses two side information bits. Since we assume similar noise characteristics for the channels from the source to all the receivers, we can compare the performances of different receivers. In that regard, among all the receivers, $R_2$ and $R_3$ perform the same and the best, followed by $R_5$. $R_1$ performs worse than $R_5$ but better than $R_4$.

Now, we will consider the same setup but with the side information being broadcast at a higher value of $\Gamma_{si}$, equal to $12$ dB. At $R_1$, there are four decoding strategies, for each of which the set  $P(\mathcal{D}^1_j), \ j \in [4]$ is determined as explained in Section \ref{mainresult1} . $P(\mathcal{D}^1_1) = 2$,  $|P(\mathcal{D}^1_2)| = 4$, $|P(\mathcal{D}^1_3)| = 8$ and $|P(\mathcal{D}^1_4)| = 6$. Hence, using Theorem \ref{thm:OptDS}, we find that an optimal decoding strategy at receiver $R_1$ is $\mathcal{D}^1_1$. At $R_2$, there are four decoding strategies, $|P(\mathcal{D}^2_1)|$ is two, $|P(\mathcal{D}^2_2)|=4$, $|P(\mathcal{D}^2_3)| = 4$, whereas for $|P(\mathcal{D}^2_4)|=6$ . Hence, an optimal decoding strategy is $\mathcal{D}^2_1$. For the receivers $R_3, R_4$ and $R_5$, there is only one decoding strategy for which  $|P(\mathcal{D}^3_1)| = 2$, $|P(\mathcal{D}^4_1)| = 8$ and $|P(\mathcal{D}^5_1)| = 6$.
So, in the high $\Gamma_{si}$ regime, the best performance at the receivers $R_1, R_2$ and $R_3$ are obtained by using the decoding strategies $\mathcal{D}_1^1$, $\mathcal{D}_1^2$, and $\mathcal{D}_1^3$, respectively. All these receivers will perform the same and the best among all the receivers. The receiver $R_5$ employing the decoding strategy $\mathcal{D}^5_1$, which has six adjacent signal pairs, will perform second-best, followed by $R_4$, performing the worst among all the receivers.

		\section{Validity of Results over Fading Channel}
	\label{fading_analysis}
	For a given ICP and a chosen index code, assume that the index-coded vector after modulation using an $M$-PSK constellation is transmitted over a fading channel. For an index-coded vector $\mathbf{y}$, the received signal at a receiver $R_i$ is $c_i = h_i \mathcal{M}(\mathbf{y}) + n_i$, where $h_i$ is the fade coefficient of the channel between the source and the receiver $R_i$, $\mathcal{M}$ denotes the mapping from the index-coded vector to a signal point in the $M$-PSK constellation, and $n_i$ is the Gaussian noise added at $R_i$. Assume that the receivers have perfect channel state information and follow the two-step process of first estimating the transmitted signal point or, equivalently, the index-coded bits and then index-decoding to estimate their desired messages. 
	
	Under this assumption, the results in Section \ref{noiseless_si} and Section \ref{noisy_si} continue to hold and the best decoding strategy at a receiver in low $\Gamma_{si}$ regime is the one which utilizes least side information bits while at high $\Gamma_{si}$ regime and noiseless side information case, the best decoding strategy continues to be the one with the minimum multiplicity of signal pairs at a distance equal to the minimum Euclidean distance of the constellation used.

	In the following section, we give simulation results validating that the probability of error performance indeed depends on the criterion given in Theorem \ref{thm:pe} and Theorem \ref{thm:noisyPe} for noiseless and noisy side information respectively.

		\section{Simulation Results}
	\label{simulationexample}
	For the noisy ICPs  in Example \ref{example-1} and Example \ref{example-2}, where the transmission of the index code is done after modulating using PSK signal sets with mapping Fig.\ref{8-psk} over AWGN broadcast channels, the probability of error performance in decoding the requested message at the receivers are simulated. For comparing the performances of different receivers, we assume that the characteristics of the Gaussian noise added at each of these receivers are the same. Let the noise be distributed as $\mathcal{N}(0, N_o)$, and energy per bit used for transmission is denoted as $E_b$. For fading channel, we are considering the perfect channel state condition with fade coefficients distribution $\mathcal{C}\mathcal{N}(0,1)$. 
	
\begin{figure}
		\begin{subfigure}{0.5\textwidth}
			\captionsetup{justification = centering}
			\captionsetup{font=small,labelfont=small}
			\includegraphics[width = \textwidth]{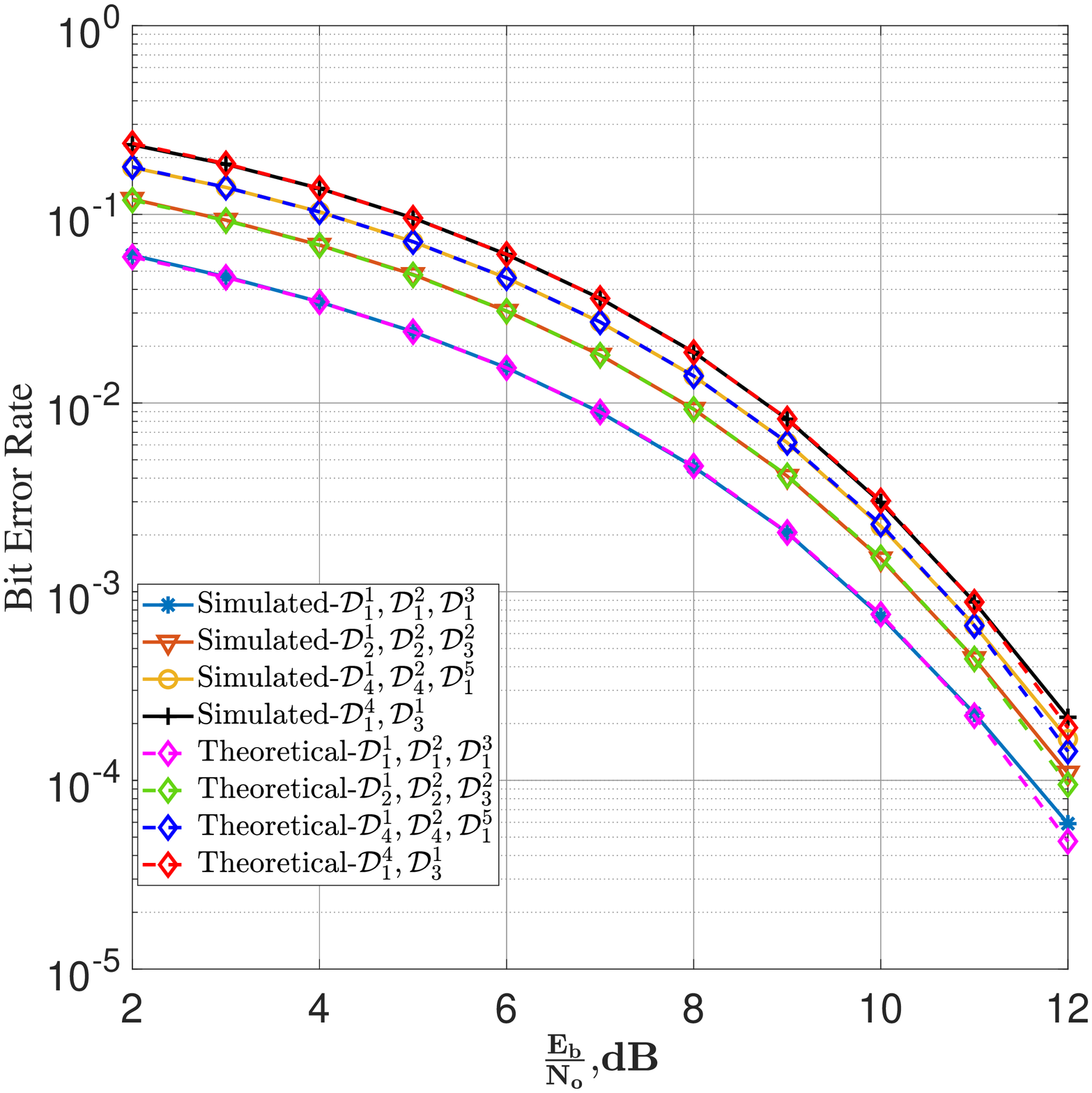}
			\caption{ Over AWGN Channel }
			\label{simulationex1_noisless_awgn}
		\end{subfigure}
		\hfill
		\begin{subfigure}{0.5\textwidth}
			\captionsetup{justification = centering}
			\captionsetup{font=small,labelfont=small}
			\includegraphics[width = \textwidth]{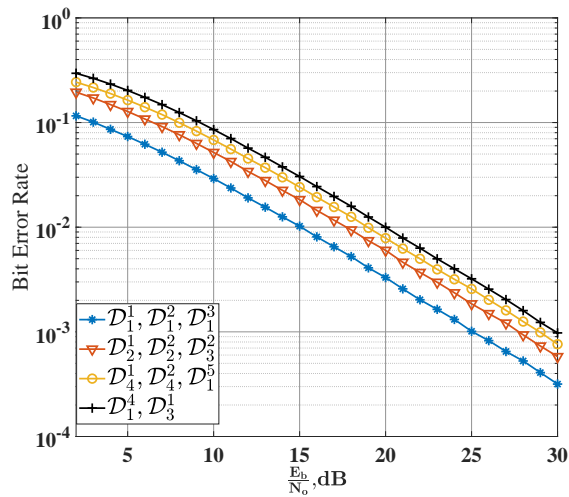}
			\caption{Over Rayleigh Fading Channel}
			\label{simulationex1_noisless_rayleigh}
		\end{subfigure}
		\caption{\textbf{Example 1}- Probability of error performances of decoding strategies at receivers with noiseless side information over noisy channels.}
	\end{figure}

	Consider the ICP in Example \ref{example-1}, for this problem, we can clearly see from Fig. \ref{simulationex1_noisless_awgn} that the best probability of error performance at $R_1$ is obtained for the decoding strategy $\mathcal{D}^1_1$, followed by $\mathcal{D}^1_2$, then $\mathcal{D}^1_4$ with $\mathcal{D}^1_3$ giving the worst performance. This ordering of the decoding strategies $\{\mathcal{D}^1_j\}_{j \in [4]}$ is the same as the increasing order of the number of elements in their corresponding sets $\{P(\mathcal{D}^1_j)\}_{j \in [4]}$. Similarly at receiver $R_2$, we have $|P(\mathcal{D}^2_1)| < |P(\mathcal{D}^2_2)| = |P(\mathcal{D}^2_3)| < |P(\mathcal{D}^2_4)|$. It can be seen that the probability of error performances of the decoding strategies at $R_2$ follow this same order in the simulation results in Fig. \ref{simulationex1_noisless_awgn}. The result in Corollary \ref{cor:Num_bits} is also validated by this simulation result, as for $R_2$, a decoding strategy employing a linear combination of two estimated index-coded bits ($\mathcal{D}^2_1$) performs better than another strategy using only one index-coded bit ($\mathcal{D}^2_2$). 
	
	Further, across different receivers, since we are assuming the same noise characteristics, the optimal decoding strategies at $R_1$ and $R_2$, as well as the only available decoding strategy at $R_3$, all give the same probability of error performance as the number of signal pairs in $P(\mathcal{D}^1_1)$, $P(\mathcal{D}^2_1)$ and $P(\mathcal{D}^3_1)$ are all equal to two. Their performances are also the best among all the receivers. $R_5$ performs second-best to it with $R_4$ performing the worst among all. From Fig. \ref{simulationex1_noisless_awgn}, we can also see that the upper bound for the probability of estimated message error achieved by a given decoding strategy in \eqref{awgnperr} is tight as it agrees very well with the simulated probability of error curves.

\begin{figure}
		\begin{subfigure}{0.5\textwidth}
			\captionsetup{justification = centering}
			\captionsetup{font=small,labelfont=small}
			\includegraphics[width = \textwidth]{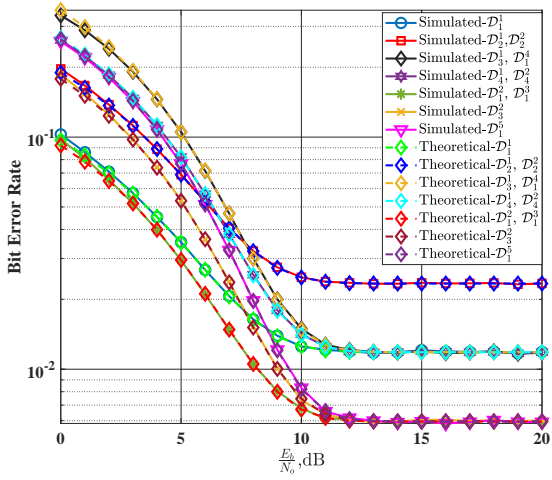}
			\caption{ At $\Gamma_{si}=5$ dB. }
			\label{simulationex1_noisy_awgn_lowsnr}
		\end{subfigure}
		\hfill
		\begin{subfigure}{0.5\textwidth}
			\captionsetup{justification = centering}
			\captionsetup{font=small,labelfont=small}
			\includegraphics[width = \textwidth]{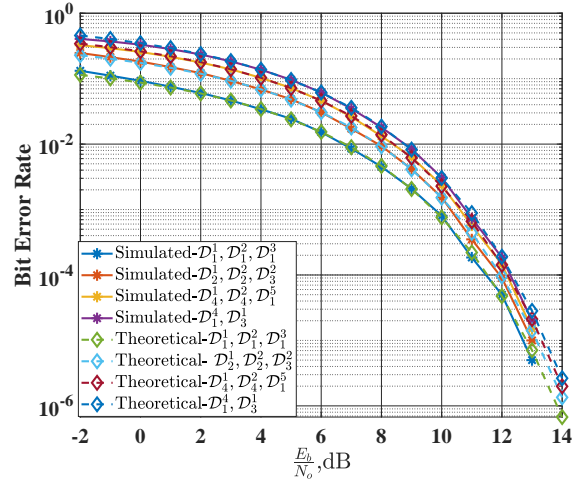}
			\caption{At $\Gamma_{si}=12$ dB.}
			\label{simulationex1_noisy_awgn_highsnr}
		\end{subfigure}
		\caption{\textbf{Example 2}- Simulated and theoretical  probability of error performances of decoding strategies at receivers with noisy side information in AWGN channel.}
		\label{simulations_noisy_awgn}
	\end{figure}

	Consider the ICP in Example \ref{example-2}, we will broadcast side information at low SNR, say $\Gamma_{si}$=5dB. We can clearly see from Fig.\ref{simulationex1_noisy_awgn_lowsnr} that the best probability of error performance for $R_1$ can be obtained for decoding strategy $\mathcal{D}_1^1$, $\mathcal{D}_3^1$ and $\mathcal{D}_4^1$ whereas $\mathcal{D}_2^1$ shows the worst performance which follows the increasing order of the number of side information bits used in each decoding strategy. Out of $\mathcal{D}_1^1$, $\mathcal{D}_3^1$ and $\mathcal{D}_4^1$ we can see that $\mathcal{D}_1^1$ performs best as $P(\mathcal{D}_1^1)$=2 which is least while that of $\mathcal{D}_3^1$ is 8. Similarly, at $R_2$, the best performance is obtained for $\mathcal{D}_1^2$, order of the number of side information used in different decoding strategies is $|S(\mathcal{D}^2_1)|=|S(\mathcal{D}^2_3)| < |S(\mathcal{D}^2_4)| < |S(\mathcal{D}^2_3)|$. In Fig.\ref{simulationex1_noisy_awgn_lowsnr}, we can see the performance of decoding strategies follow the same order. For $R_3, R_4$, and $R_5$, there is only one decoding strategy. 
	
\begin{figure}
		\begin{subfigure}{0.5\textwidth}
			\captionsetup{justification = centering}
			\captionsetup{font=small,labelfont=small}
			\includegraphics[width = \textwidth]{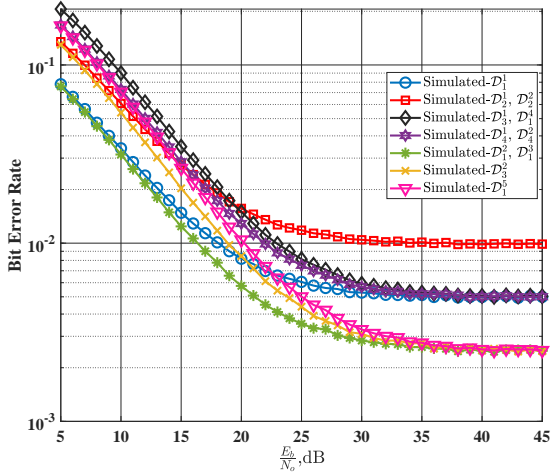}
			\caption{ At $\Gamma_{si}=20$ dB. }
			\label{simulationex1_noisy_rayleigh_lowsnr}
		\end{subfigure}
		\hfill
		\begin{subfigure}{0.5\textwidth}
			\captionsetup{justification = centering}
			\captionsetup{font=small,labelfont=small}
			\includegraphics[width = \textwidth]{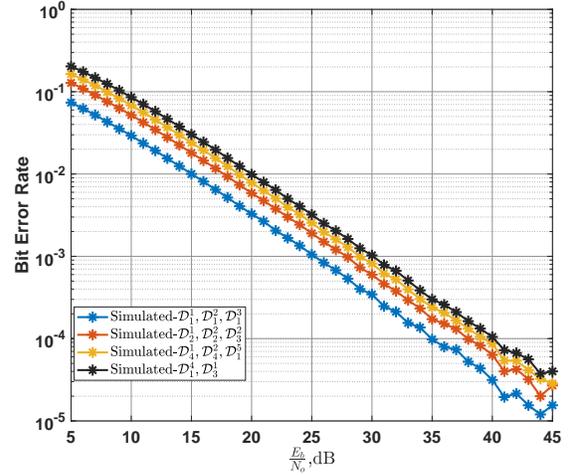}
			\caption{At $\Gamma_{si}=45$ dB.}
			\label{simulationex1_noisy_rayleigh_highsnr}
		\end{subfigure}
		\caption{\textbf{Example 2}- Simulated probability of error performances of decoding strategies at receivers with noisy side information in Rayleigh fading channel.}
		\label{simulations_noisy_rayleigh}
	\end{figure}

	Among the receivers, $R_2$ and $R_3$ perform the same and the best, followed by $R_5$, which uses one side information bit, followed by $R_1$ and then $R_4$ which performs the worst. We have discussed the performances of the receivers over the AWGN channel, but the same order can be observed over a fading channel also at $\Gamma_{si}$=20dB, as shown in Fig.\ref{simulationex1_noisy_rayleigh_lowsnr}.

	Now considering the same set-up with $\Gamma_{si}$=12 dB.	For this problem, we can clearly see from Fig. \ref{simulationex1_noisy_awgn_highsnr} that the best probability of error performance at $R_1$ is obtained for the decoding strategy $\mathcal{D}^1_1$, followed by $\mathcal{D}^1_2$, then $\mathcal{D}^1_4$ with $\mathcal{D}^1_3$ giving the worst performance. This ordering of the decoding strategies $\{\mathcal{D}^1_j\}_{j \in [4]}$ is the same as the increasing order of the number of elements in their corresponding sets $\{P(\mathcal{D}^1_j)\}_{j \in [4]}$. Similarly at receiver $R_2$, we have $|P(\mathcal{D}^2_1)| < |P(\mathcal{D}^2_2)| = |P(\mathcal{D}^2_3)| < |P(\mathcal{D}^2_4)|$. It can be seen that the probability of error performances of the decoding strategies at $R_2$ follow this same order in the simulation results in Fig. \ref{simulationex1_noisy_awgn_highsnr}. 
	
	Further, across different receivers, since we are assuming the same noise characteristics, optimal decoding strategies at $R_1$ and $R_2$, as well as the only available decoding strategy at $R_3$, all give the same probability of error performance as the number of signal-pairs in $P(\mathcal{D}^1_1)$, $P(\mathcal{D}^2_1)$ and $P(\mathcal{D}^3_1)$ are all equal to two. Their performances are also the best among all the receivers. $R_5$ performs second-best to it with $R_4$ performing the worst among all. Over a fading channel, under perfect channel state assumption, in a high SNR regime, say $\Gamma_{si}=45$dB, it can be seen in Fig.\ref{simulationex1_noisy_rayleigh_highsnr} that the order of performances of the decoding strategies is the same as that over an AWGN channel.
	

	\section{Conclusion}
	\label{conclusion}
	For a given ICP with unprioritized receivers, we proved that when multi-level modulation schemes are used for transmitting the index-coded bits over a noisy channel, then minimizing the number of transmission does not assure the best probability of error performance. Further when we consider the noisy side information of receivers, the performance depends on signal-to-noise ratio ($\Gamma_{si}$) at which it is broadcast. We derived the criterion for selecting an optimal decoding strategy at low and high $\Gamma_{si}$ regime and have also proposed an algorithm for determining optimal decoding strategy at a given receiver. 
	
%

	\end{document}